# High power density energy harvesting devices based on the anomalous Nernst effect of Co/Pt magnetic multilayers

*Guillermo Lopez-Polin*[*,1], *Hugo Aramberri*[2], *Jorge Marques-Marchan*[1], *Benjamin I. Weintrub*[3], *Kirill I. Bolotin*[3], *Jorge I. Cerdá*[1], *Agustina Asenjo*[*,1]

[1]Instituto de Ciencia de Materiales de Madrid (ICMM-CSIC), 28049 Madrid, Spain

[2]Materials Research and Technology Department, Luxembourg Institute of Science and Technology (LIST), L-4362 Esch/Alzette, Luxembourg

[3]Department of Physics, Freie University Berlin 14195 Berlin, Germany

AUTHOR INFORMATION

**Corresponding Author**

* Guillermo Lopez-Polin. e-mail: guillermo.lopez-polin@uam.es

*Agustina Asenjo: e-mail: aasenjo@icmm.csic.es

The anomalous Nernst effect (ANE) is a thermomagnetic phenomenon with potential applications in thermal energy harvesting. While many recent works studied the approaches to increase the ANE coefficient of materials, relatively little effort was devoted to increasing the power supplied by the effect. Here we demonstrate a nanofabricated device with record power density generated by the ANE. To accomplish this, we fabricate micrometer-sized devices in which the thermal gradient is three orders of magnitude higher than conventional macroscopic devices. In addition, we use Co/Pt multilayers, a system characterized by a high ANE thermopower (~1 µV/K), low electrical resistivity, and perpendicular magnetic anisotropy. These innovations allow us to obtain power densities of around 13±2 W/cm$^3$. We believe that this design may find uses in harvesting wasted energy in e.g. electronic devices.

**Introduction**

Thermoelectric generators, traditionally based on the Seebeck effect[1], are devices capable of harvesting energy from any source of wasted thermal energy, such as the heat emitted by the human body or by the Joule effect of different electronic devices. More than half of the considered waste heat streams arise at a low temperature range, below 200 °C, of which the microelectronics take an important share[2]. This paper explores the opportunity of using thermoelectric effects for converting wasted heat from high-performance integrated circuits, such as microprocessors, into electric energy. During the last decade, thermoelectric generators based on the anomalous Nernst effect (ANE) have been proposed for their potential to achieve a better efficiency[3]. The ANE occurs in the presence of a thermal gradient perpendicular to a magnetic field, resulting in an electric field perpendicular to both[3, 4]. The effect is anomalous when it is produced by the intrinsic magnetization of the sample instead of an external magnetic field. The transverse nature of ANE (in contrast with the Seebeck effect) allows for lateral configurations with simpler design, higher performance and much lower production cost. Recently, materials with increasing ANE coefficients and decreasing costs have been found[5-7]. However, ANE generators still have some disadvantages compared to those based on the Seebeck effect. Firstly, the maximum ANE thermopower measured so far [7, 8] is still two orders of magnitude lower than the maximum Seebeck coefficient found in an element[9]. Moreover, while the Seebeck effect is proportional to the gradient of temperature, the ANE depends also on the component of the magnetization perpendicular to the thermal gradient ($\vec{E}_{ANE} \propto \vec{m} \times \vec{\nabla} T$). Therefore, to have the magnetization in a well-defined direction becomes crucial for the enhancement of this effect. Magnetic multilayers showing high perpendicular magnetic anisotropy (PMA) have been studied for long time because of their critical role in the development of magnetic recording[10], high-density non-volatile memories[11], interface-induced phenomena[12], emergent spin-electronic technology[13], and the discovery of the magnetic skyrmions[14]. But materials with PMA are also very advantageous for maximizing the ANE voltage owing to their high remanent magnetization along the out-of-plane direction[15-17].

The ANE can be viewed as the thermal analog of the anomalous Hall effect (AHE) in magnetic materials[7]. The AHE causes the appearance of a voltage difference across a ferromagnetic electrical conductor perpendicular to an electric current flowing through the material. In the ANE the driving current is caused by the Seebeck effect (i.e. parallel to the thermal gradient), resulting in an electric field transverse to the thermal gradient. The AHE is a phenomenon where the transport properties of the spin-polarized electrons are governed by the spin-orbit coupling (SOC) which couples the orbital and spin degrees of freedom of the electron[18]. Consequently, SOC plays a crucial role in both AHE and ANE[7]. Both effects are connected to the electrical conductivity through the Mott relations[19]. Most of the materials with high ANE also show relatively large remanent magnetization. In the last few years, however, it was shown that a substantial contribution to the ANE originates from the Berry curvature close to the Fermi level[20], and hence is strongly affected by topological band crossings (like Weyl nodes) that act as sources or sinks of the Berry curvature (and are themselves affected by the magnetization). Following this realization, some materials with large ANE and low magnetization have been reported[21]. This contribution is often called "intrinsic", as opposed to other so-called extrinsic mechanisms like magnon drag[7, 22], phonon drag[23] or skew forces due to large spin-orbit coupling in ferromagnetic metals[24, 25], which have also been proposed to yield large contributions to the ANE.

Most of the works published about the ANE coefficient of different materials focus on the ANE voltage but no attention has been paid to the maximum power supplied by the ANE effect. Power supplied by the ANE will play a major role in determining the feasibility of the material for energy harvesting applications. In this work, we study the ANE response of Co/Pt multilayers with high PMA, in which the magnetization of the system is well oriented perpendicular to the layers. In addition, the miniaturization of the heat source produces the thermal decay to occur in a reduced space, thus enhancing the thermal gradient (in K/m). Therefore, from miniaturized devices we expect to obtain higher voltages and current densities. Accordingly, we evaluate the power supplied by Co/Pt devices with micrometer lateral dimensions, obtaining record power densities supplied by our devices. Our data suggest that thermoelectric devices based on the Nernst effect can be very effective to harvest wasted thermal energy from nano- or microscopic heating sources like microelectronic devices.

**Experimental Section**

We first explore the ANE response in macroscopic devices based on Co/Pt multilayer samples. In order to evaluate the influence of the Co and Pt thicknesses, we fabricate several samples on SiNx /Si substrates with different Co and Pt thicknesses and a fixed number of repetitions (namely 10 repetitions). The ANE response of these multilayers is evaluated by using the setup for characterizing the ANE of macroscopic devices described in Supporting Information (SI) 1.a. We use the configuration with in-plane temperature gradient and out-of-plane magnetic fields to avoid contributions from the spin Seebeck effect[17]. In addition, this setup allows us to determine the temperature gradient more precisely in order to correctly characterize the ANE thermopower. The maximum ANE response in saturation was found for the multilayers with 0.5nm of Co and 1.5nm of Pt, which show high PMA as deduced from the magnetic hysteresis loops (SI 1.a) measured by vibrating sample magnetometer (VSM). Hereafter we focus on the characterization of the devices based on these $[Co_{0.5nm}Pt_{1.5nm}]_{10}$ multilayers (TEM characterization in SI1b).

In order to achieve higher thermal gradients and optimize the voltage per area, we fabricate microscopic devices as illustrated in Figure 1.a (more details in SI 1.c). The 100x5 µm² stripe at the center (dark blue colored in the zoomed of Figure 1.c) is the Co/Pt multilayer. The rest of the patterns were made of Pt: the bar on the left (orange) acts as a heater, and the bar on the right (light blue) works as a four-terminal thermal sensor. The high PMA of the selected multilayer [Co$_{0.5nm}$Pt$_{1.5nm}$]$_{10}$ is revealed from in-plane and out-of-plane hysteresis loops represented in Figure 1.b, showing high out-of-plane remanent magnetization (close to the saturation value) and low in-plane remanence. In addition, hysteresis loops show a remanence of the magnetization of ~100% along the perpendicular direction with no applied field, in contrast with other previously reported materials with a high ANE coefficient but very low remanence[5,6, 26]. The hysteresis loops were measured by VSM on 5x5mm² samples grown in parallel with the multilayer structure of the microscopic devices. In the following, all the reported experiments are performed on the micro-fabricated devices. We measure the magnetic state of the multilayer by magnetic force microscopy (MFM) using amplitude modulation mode and a low magnetic moment tip (SI6). Figure 1.d shows the topography, and the profile shows the geometry of the device: the thickness of the multilayers, the heater and the thermometer, as well as the distance between them. Figure 1.e is the corresponding MFM image obtained in a demagnetized state at a retrace distance of 30 nm. The phase-locked loop feedback is activated to keep the phase constant, and thus the magnetic signal is in the frequency shift channel [27]. The variation of the resonance frequency at this distance mostly depends on long-range conservative forces (i.e. magnetic and electrostatic forces). The MFM images in this state exhibit stripe domains. From these images we can extract the normalized remanent magnetization of the sample in the perpendicular direction, M$_{RZ}$, from[28]:

$$M_{RZ} = \frac{A\uparrow - A\downarrow}{A\uparrow + A\downarrow} \qquad (1)$$

where $A\uparrow$ is the area of the domains pointing up and $A\downarrow$ is the area of the domains pointing down. We assume a constant magnetization across the thickness of the sample. In order to characterize the magnetization in situ and relate the ANE response with the domain structure of the multilayer, we measure the ANE voltage while simultaneously obtaining MFM images and vary the magnetization of the sample with a perpendicular field created by an electromagnet below the sample[29].

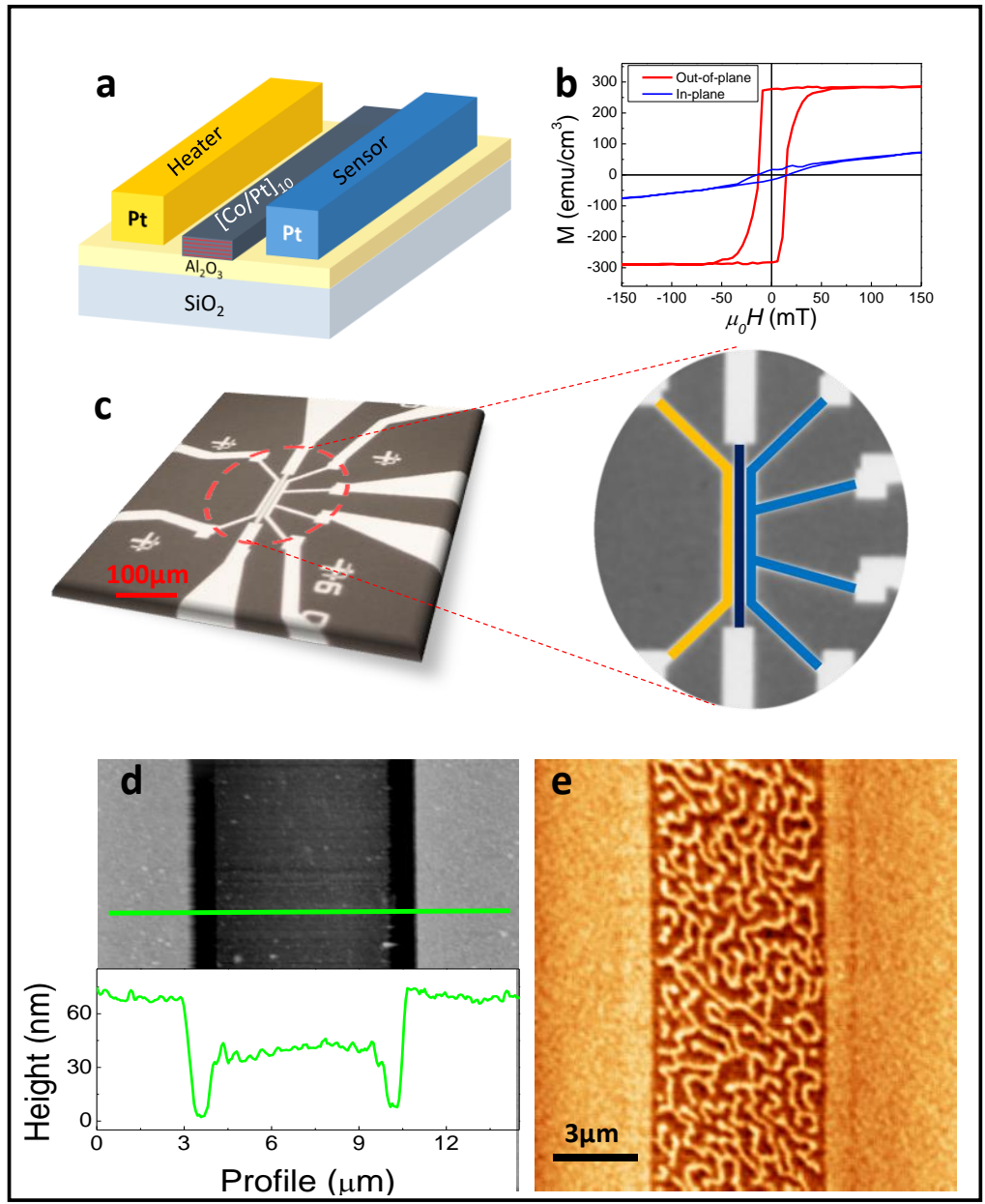

*Figure 1* a) Cartoon showing a schematic of the device. b) In-plane (blue) and out-of-plane (red) hysteresis loops. c) Optical image of the device and zoom with colored regions corresponding to the stripe (dark blue), the heater (yellow) and the sensor (light blue). d) AFM topography of the heater, Co/Pt multilayer and sensor. Green line corresponds to the line of profile shown below. e) MFM image of the magnetic stripe in a demagnetized state.

A key issue to accurately evaluate the ANE thermopower and coefficient of the devices is to extract the temperature gradient from the average temperature of the Pt sensor, which is the magnitude measurable in our experiments. We carry out COMSOL Multiphysics simulations[30] to extract the temperature gradient in the Co/Pt structure from the average temperature of the platinum sensor. A more detailed description of the finite-element simulations and the method for determining the temperature gradient can be found in SI2. We find that the mean thermal gradient across the multilayer structure has a linear dependence with the average temperature of the sensor of about 0.08 K/µm per K measured in the resistance. This value will be used to calculate the thermal gradient in the Co/Pt element based on the experimental values of the Pt sensor. The relationship between the resistance of the Pt sensor and the temperature is determined experimentally. A linear dependence of 2700 ppm/K is obtained, much lower than the value for bulk Pt[31] (SI 3).

**Results and Discussion**

The characterization of the device has been performed using different strategies: a complete map of the ANE voltage versus temperature and magnetic field, and individual curves of ANE voltage either versus temperature (in remanence) or versus magnetic field (at a fixed temperature). To experimentally control the thermal gradient variation, we sweep the current passing through the heater. Similarly, the in situ out-of-plane magnetic field can be modified continuously[29]. Figure 2.a shows a map of the ANE voltage ($V_{ANE}$) as a function of the thermal gradient across the multilayer (varying the current from negative to positive values along the fast scan, X axis) and the out-of-plane magnetic field (slow scan, Y axis). Firstly, we measure $V_{ANE}$ versus temperature gradient at the maximum negative field (-45 mT) that we can maintain in the system for long periods of time without overheating the coil (bottom of the map in Fig. 2.a). Then, we apply increasing magnetic fields from negative to positive values while $V_{ANE}$ versus temperature curves are measured at each field. As the heater increases the temperature with the current, regardless of the direction, the Nernst voltage versus the current presents a V shape. This allows us to discard possible current leaks from the heater or the sensor to the structure. Figure 2.b shows the hysteresis loop of $V_{ANE}$ at a constant thermal gradient of ~2.5 K/µm, sweeping the field from -60 mT to 60 mT (which is close to the saturation field with > 90% alignment). This data is in good agreement with Figure 2.a since similar behavior can be found by plotting a vertical profile through the 3D map. Notice that the hysteretic behavior is similar to the magnetization vs field plot, evidencing the thermomagnetic origin of the effect.

Besides, the in situ $V_{ANE}$ signal is measured as a function of the temperature gradient in the remanence state. The $V_{ANE}$ exhibits a lineal dependence with temperature (Figure 2.c, additional information in SI4 and SI5). Similar information can be obtained from the horizontal profiles in Figure 2.a, which corresponds to a $V_{ANE}$ curve versus temperature under an applied magnetic field. The MFM images obtained in remanence, after applying a saturating out-of-plane magnetic field ex situ, present a uniform contrast, which corresponds to a remanence of ~100%. (SI6). Finally, we repeat these curves (ANE response versus the current applied to the heater) for the remanent states reached after applying different magnetic fields. The magnetization of the sample is evaluated by using the MFM images. The ANE thermopower ($S_{yx}=S_{ANE}$) of the sample was extracted by multiplying the slope of each curve (total voltage divided by the temperature difference across the whole multilayer in µV/K) by the geometrical factor of the magnetic stripe (length divided by width). Figure 2.d shows the $S_{ANE}$ as a function of the remanent magnetization of the sample. As expected, the $S_{ANE}$ exhibits a linear trend with the magnetization, and reaches the same absolute value at $M/M_S = 1$ and at $M/M_S = -1$. When the sample is saturated, we obtain a value of 0.95 µV/K, very similar to the ANE obtained on the macroscopic device corresponding to the same multilayered sample ($Co_{0.5}Pt_{1.5}$). From the slope we can extract the Nernst coefficient, that is defined by:

$$N = \frac{V_{ANE}}{\mu_0 M \overline{\Delta T}} \frac{w}{l} = \frac{S_{ANE}}{\mu_0 M} \qquad (2)$$

where $\mu_0$ is the vacuum permeability, M is the maximum magnetization of the structure, $\overline{\Delta T}$ is the average thermal gradient, w is the width and l the length of the multilayer bar. The Nernst coefficient obtained for this multilayer is N=2.3 µV/(T·K).

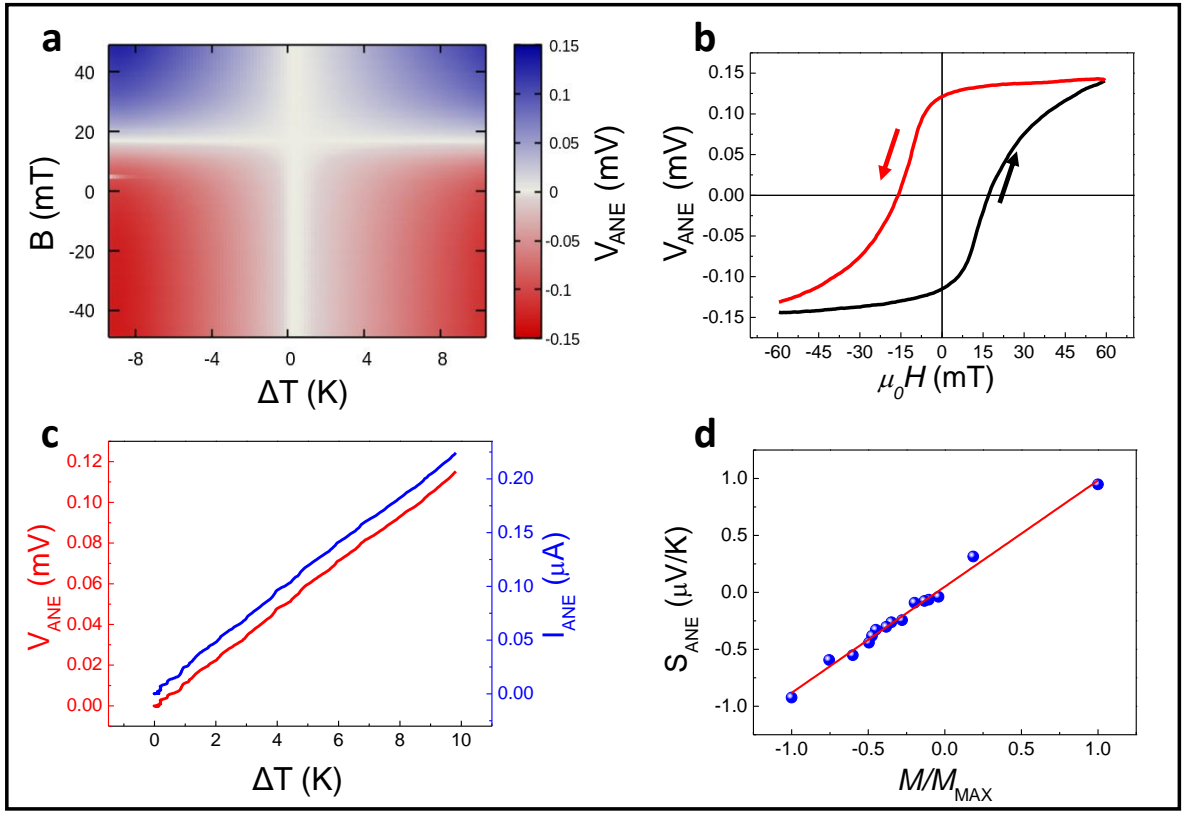

*Figure 2 a) Map of the Nernst voltage as a function of the thermal gradient (x axis) and the applied magnetic field (y axis). Notice that the negative sign of the thermal gradient is only indicating the direction of the current passing through the heater. b) Hysteresis loop of the Nernst voltage vs the*

*applied field. c) Nernst voltage (red) and generated current (blue) as a function of thermal gradient. d) ANE thermopower as a function of the magnetization of the sample.*

In order to make an ANE device with high energy-harvesting potential, it is also crucial to evaluate the maximum power that the device can provide. We measured the current supplied by the multilayer as a function of the thermal gradient by grounding one of the sides of the device through a current-to-voltage converter while applying a thermal gradient. Notice that the larger the thermal gradient, the higher the voltage (and current) obtained (see Figure 2.c). However, the maximum performance of the device is limited by the maximum temperature admissible until the magnetic moment starts to decrease due to thermal fluctuations (SI4). In this case, the maximum voltage and current are ~0.3 mV and ~0.6 µA with a total thermal gradient of ~4 K/µm. For comparison, the maximum gradient of temperatures achieved on the macroscopic devices is around 0.003 K/µm. It is important to consider that the measurement of the current was performed with two terminals. Thus, the current is not only dictated by the resistance of the structure but also the contact resistance. In fact, the resistance of the multilayer measured in two-contacts was about 500 Ω, which is just the voltage divided by the obtained current. Therefore, the measured value (0.6 µA) is just a lower limit of the maximum current that can be supplied by the device, but it could be higher if the contact resistance is reduced. The maximum power supplied by the device is $P = IV$, which gives approximately 180 pW. The power scales linearly with every dimension of the structure: voltage scales linearly with the length of the device while current remains constant, and the current increases proportionally to the width and thickness of the multilayer with no variation of the voltage. At the largest temperature gradient, the maximum voltage per length obtained in remanence is 30 mV/cm and the maximum current per section area is 480 A/cm². From three different samples we obtained a maximum power density of 13±2 W/cm$^3$. Note that Seebeck generators usually give power densities in the order of tens of mW/cm$^3$ [32].

The high ANE voltages obtained from the devices due to the high thermal gradients achieved and the high Nernst coefficient of the multilayers make it possible to detect small variations of the magnetization of the device. To quantify the limit of the sensitivity of the device we have induced slight variations in the total magnetization of the sample. With the tip stray field, antiparallel to the sample magnetization, it is possible to 'write' domains inducing local changes in the magnetic state[15] just by approaching the MFM tip to certain regions of the sample. With just one sweep we inverted around 40% of the scanned area, around 0.25 µm x 5 µm (Figure 3.a and 3.b). This slight variation of the total magnetization of the sample, around ~0.1%, can be detected by comparing the $V_{ANE}$ before and after the writing process as shown in Figure 3.c. The expected linear decay of the |$S_{ANE}$| versus the magnetic moment of the sample (proportional to the volume of the sample pointing in down direction) is shown in Figure 3.d. The sensitivity of the device is enough to detect human heat in ambient conditions (see SI9).

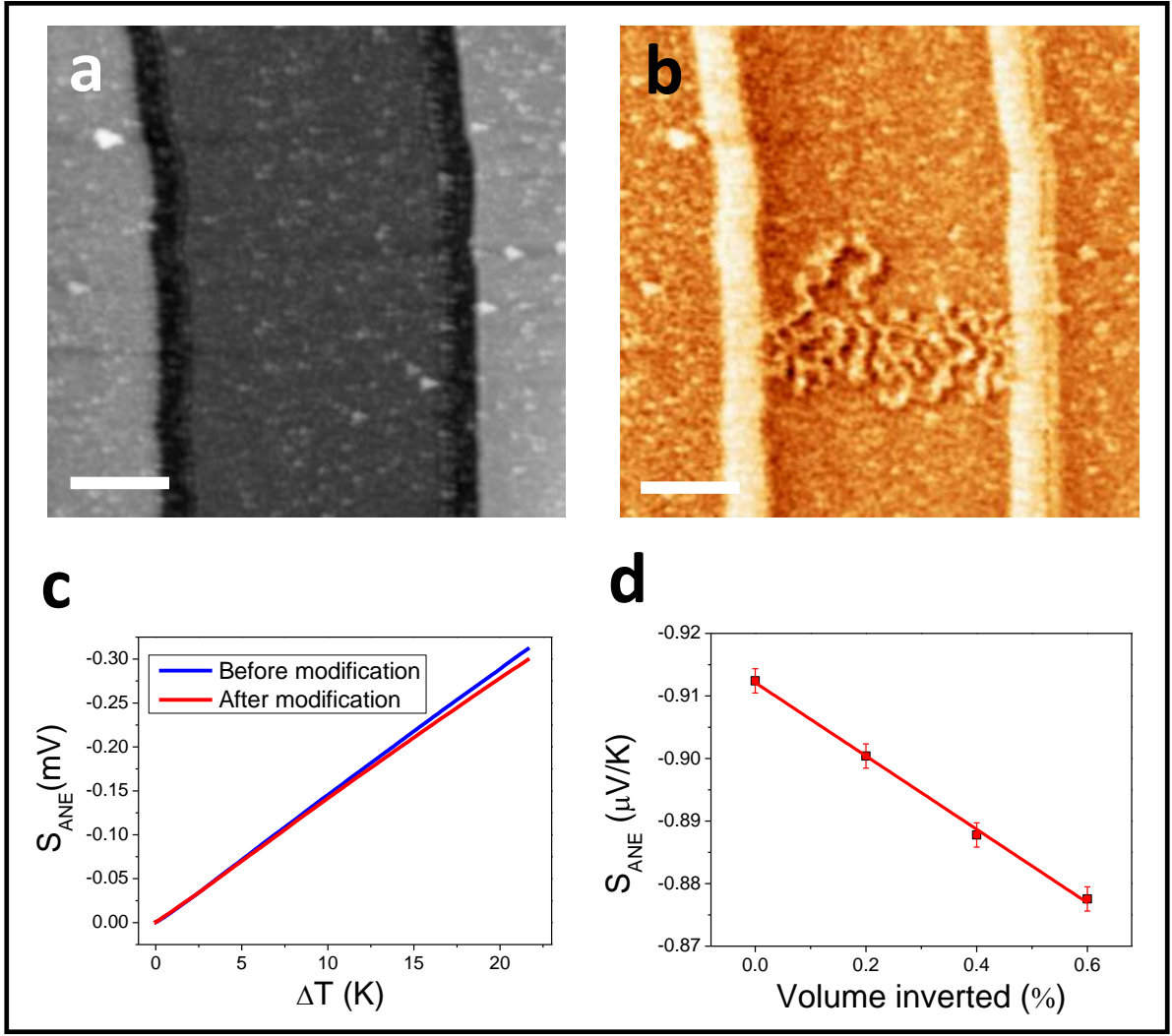

*Figure 3* a) Topography and b) MFM image of the device after modification, c) ANE voltage (before and after modification) versus temperature, d) variation of the ANE thermopower with the % of volume inverted with the tip.

The ANE of other materials with PMA have also been measured by other groups[15-17, 33]. While some of them present an ANE thermopower lower than that reported here (e.g. $L1_0$-Ordered epitaxial FePt thin films[16] and multilayers of MgO/$Co_2$MnGa/Pd[33]), others portray a Nernst thermopower almost twice as large (e.g. IrMn/CoFeB/MgO[17]). However, the resistivity of IrMn/CoFeB/MgO should be very high because most of the layers show very poor conductivity: MgO is an insulator and CoFeB presents high resistivity[34]. Therefore, maximum power provided by these multilayers should be much lower than that reported in this work. In a previous work on MnBi[7], a large ANE was attributed to an extrinsic magnon-drag effect, which is most likely to occur in ferromagnetic samples with large SOC, as the one discussed here. Besides, the ANE thermopower of Fe/Pt multilayers was reported to be ~1 µV/K[35], very similar to the value measured in this work. The advantage of our system is its high PMA, which simplifies the design of a device, allowing to easily align the direction of the thermal gradient perpendicular to the magnetization to obtain the maximum ANE voltage. The high ANE measured on Fe/Pt multilayers was attributed to an unconventional interface-induced enhancement thermoelectric conversion in the Fe films, not to the proximity effect-induced magnetism on the Pt by the Fe layers. Also, $UCo_{0.8}Ru_{0.2}Al$ was reported to have a colossal ANE of 23 µV K$^{-1}$ and a large ANE conductivity of 15 A m$^{-1}$ K$^{-1}$ at ~40 K [8]. However, above 60 K the material is not ferromagnetic and does not present ANE. The authors claim that to enhance the Berry curvature contribution to the ANE, in addition to a large spin-orbit coupling (SOC) and a strong electronic correlation, a kagome lattice structure can be of help. Theoretical calculations show that there are several Weyl nodes close to the Fermi level on this material. In another work on $YMnBi_2$[24], the large ANE measured in the samples could not be explained from first-principles based calculations (which typically only account for the intrinsic contribution) of the Nernst thermopower obtained with the Mott relation. Scattering (unavoidable in real samples, particularly in those with large SOC) induces skew forces and side jump on the spin-polarized charge carriers[24]. Moreover, the net ANE can have contributions of all the mentioned origins (including the Berry curvature). It is even possible that none of them is predominant. However, testing all these possibilities falls beyond the reach of the present work.

To put into perspective our results, Figure 4 shows the values of anomalous Nernst thermopowers for several works in the literature. The shaded region is that of $|S_{ij}| = |Q_s|\mu_0 M$ with $|Q_s|$ ranging from 0.05 to 0.1 µV K$^{-1}$ T$^{-1}$, where most of the reported ANE thermopowers lie. The Co/Pt multilayers here studied fall outside this region, like Mn$_{3+x}$Sn$_{1-x}$ or Co$_2$MnGa, which have been shown to display a large intrinsic ANE.

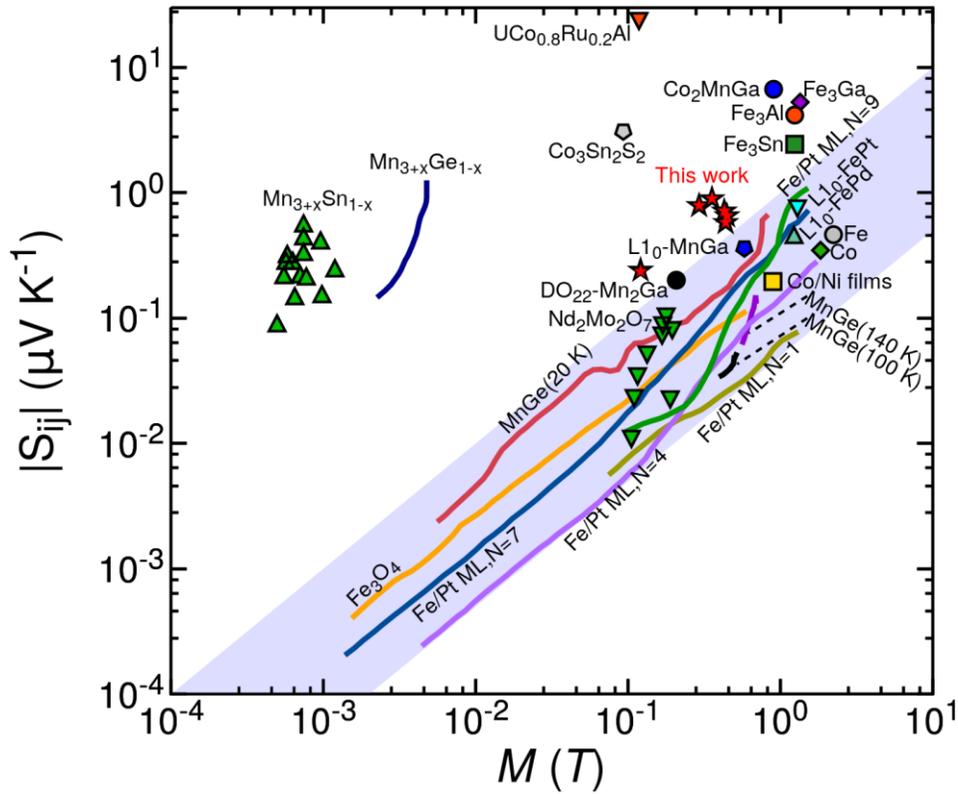

*Figure 4 Anomalous Nernst thermopower as a function of spontaneous magnetization for several materials reported in the literature. We include data for Fe$_3$O$_4$ (orange line)[19]; Mn$_{3+x}$Sn$_{1-x}$ (green upwards triangles)[21]; Mn$_{3+x}$Ge$_{1-x}$ (dark blue line)[36]; Co$_2$MnGa (blue circle)[26]; Co$_3$Sn$_2$S$_2$ (grey*

*pentagon)³⁷; UCo₀.₈Ru₀.₂Al (orange downwards triangle)⁸; Fe₃Ga (purple diamond)⁶; Fe₃Al (orange circle)⁶; Fe₃Sn (green square)³⁸; Fe/Pt multilayers with N interfaces³⁵ with N=1 (ochre line), N=4 (light purple line), N=7 (blue line) and N=9 (green line); MnGe³⁹ at T=20 K (red line), T=100 K (black lines) and T=140 K (dark purple line); Nd₂Mo₂O₇ (green downwards triangles), Fe (grey circle) and Co (green diamond)⁴⁰; L₁₀-FePd (turquoise upwards triangle), L₁₀-FePt (cyan downwards triangle), L₁₀-MnGa (blue pentagon), D₀₂₂-Mn₂Ga (black circle)⁴¹, Co/Ni films (yellow square)⁴¹ and the Co/Pt multilayers studied in this work (red stars). The shaded region indicates the linear relation for conventional ferromagnetic metals.*

In order to elucidate the origin of the large ANE in our Co/Pt multilayers, we performed AHE measurements [42] of the [Co$_{0.5nm}$Pt$_{1.5nm}$]$_{10}$ multilayers and carried out first-principles calculations (details in SI 7). To this end we grow a Hall bar of Co/Pt. We measured the longitudinal ($\rho_{XX}$) and transversal ($\rho_{XY}$) resistivity at room temperature (300K) of the multilayer as a function of an external field (Figure 5 and SI 9). We observed that the $\rho_{XY}$ vs. field (Figure 5) has a similar shape to the magnetization vs field curve. The AHE resistivity saturates at $\rho_{XY}$~1 µΩ·cm, and no increase is detectable for higher fields, which means that the anomalous contribution to the AHE is much larger than that of the ordinary Hall effect. On the other hand, the longitudinal resistivity (SI8) show no variations with the field and remains constant at $\rho_{XX}$~50 µΩ·cm. This data is in good agreement with other works determining the longitudinal and Hall resistivity of Co/Pt multilayers [43, 44]. While the longitudinal resistivity falls at the limit between the dirty (or bad metal) regime and the intrinsic regime, the very low Hall resistivity falls well beyond the universal scaling law for anomalous transport for bulk systems[42]. Previous works [43, 44] on the origin of the anomalous Hall transport on these multilayers point at interface scattering as the most important contribution to the AHE, which also modifies the relationship between ρ$_{XX}$ and ρ$_{XY}$, and explains the deviation of the multilayers from the universal scaling law of the transversal and longitudinal conductivity for bulk materials.

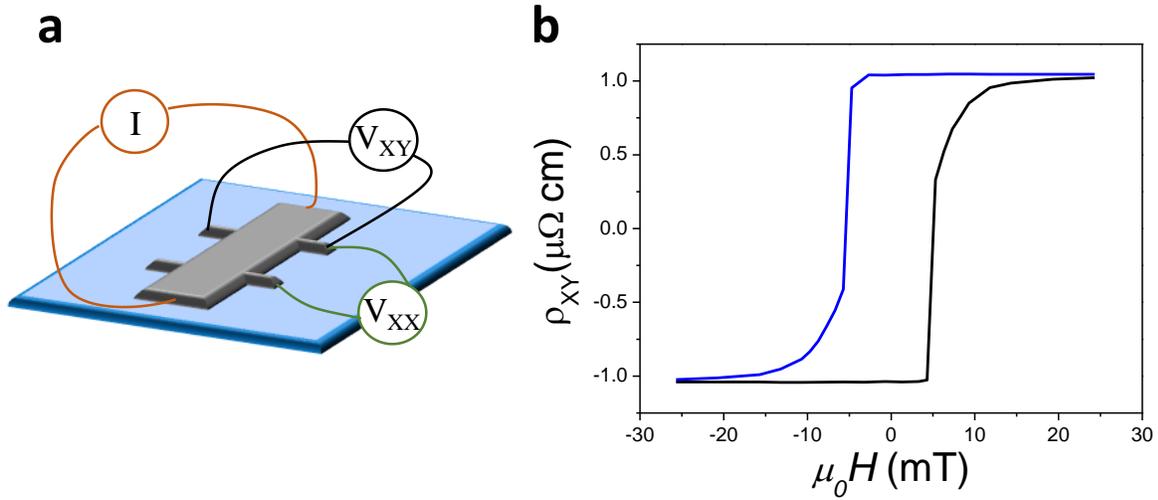

*Figure 5 a) Cartoon showing the setup used to measure the AHE of the Co/Pt multilayers. b) AHE resistivity vs applied field of the Co/Pt multilayer. The measurements are performed at RT with the magnetic field applied perpendicular to the sample. Black line is measured from negative to positive fields and blue line is measured from positive to negative fields.*

From the first-principles calculations we obtain the (intrinsic) thermoelectric conductivity tensor ($\alpha_{ij}$) of Co/Pt heterostructures (see SI7). We explore different thicknesses of the Pt layer (from 2 to 5 monolayers), while the Co layer is fixed (to 2 monolayers). The results, shown in Figure S7, indicate that the intrinsic ANE would change sign with the addition of each Pt monolayer. From the transport measurements we can estimate the experimental thermoelectric conductivity of the Co$_{0.5nm}$Pt$_{1.5nm}$ multilayer through $\alpha_{xy} = \frac{S_{xy}}{\rho_{xx}} + \frac{S_{xx}\rho_{xy}}{\rho_{xx}^2}$, where $\rho_{xx}$ is the longitudinal resistivity, $\rho_{xy}$ is the Hall resistivity and S$_{xx}$ is the Seebeck coefficient[45]. We obtain $\alpha_{xy} = 2.5$ A K$^{-1}$ m$^{-1}$, relatively far from the computed value for the 2Co/5Pt heterostructure (1.2 A K$^{-1}$ m$^{-1}$), whose layer thicknesses are the closest to the experimental ones among those calculated. More importantly, the ANE contribution to the experimental thermoelectric conductivity dominates over that of the AHE (the former is more than two times larger than the latter), and can be expected to dominate at other Pt thicknesses (note that the $\rho_{xy}$ has been shown to vary only modestly with Pt thickness[46]). Since the measured ANE thermopowers vary only mildly with Pt thickness, we anticipate that the experimental thermoelectric conductivities will also vary only slightly with Pt thicknesses. This is at odds with the computed (intrinsic) thermoelectric conductivities, which depend very strongly on the thickness. (Note that in other systems the first-principles based calculations of the intrinsic ANE also fail to fully explain the origin of large Nernst thermopower[24]). Hence, this indicates again that the origin of the main contribution to the ANE of Co/Pt multilayers is not intrinsic. Our simulations thus point at extrinsic effects as the main contribution to the large ANE in this system.

**Conclusions**

Our results show that Co/Pt multilayers present high ANE coefficient and thermopower. Materials with large PMA are ideal candidates for the development of ANE devices due to their well-defined magnetization direction that allows to optimize and improve their performance as energy harvesters. Moreover, we propose to enhance the power densities of the devices by miniaturization to the micro or nanoscale. We obtained power densities of 13±2 W/cm$^3$. Thus, usable powers could be obtained by multilayers thicker than that presented here, and by parallelizing several devices.

In contrast to the insulators that present high ANE thermopower, the advantage of multi-layered metallic systems is their low electrical resistance that allows to achieve high current densities. In addition, the technical growth requirements make this system affordable and easily achievable.

To elucidate the origin of the high ANE thermopower on the multilayers we present theoretical simulations and AHE measurements. Theoretical simulations suggest that the origin of the high ANE is not intrinsic. The longitudinal and AHE resistivity is similar to that reported in previous works, where interface scattering is found to be the main contribution to the AHE in similar multilayers with strong PMA. Therefore, we attribute the high ANE of Co/Pt multilayers also to the interface scattering.

In summary, the ANE exhibited by the FM/HM multilayers can be a novel approach to energy sustainability due to its potential to produce electric harvesting devices with large power densities. One additional benefit of this system is its ability to develop functional devices on a variety of substrates, including flexible ones. This could allow to fabricate wearable thermoelectric generators, which would be useful to feed wearable electronic devices for medical applications or related with the Internet of Things.

ASSOCIATED CONTENT

**Supporting Information**. Device fabrication. Atomic structure of the multilayers as measured by TEM. COMSOL simulations. Calibration of the dependence of the resistivity of the platinum bar with temperature. MFM images of the microscopic devices. Maximum temperature gradient and ANE voltage in remanence. $V_{ANE}$ vs $\Delta T$ curves from experimental data. First-principles calculations. Anomalous Hall effect measurements. Sensitivity of macroscopic devices.

ACKNOWLEDGMENT

This work is supported by the Spanish Ministry of Science and Innovation through the projects PID2019-108075RB-C31 and MCIN/FEDER RTI2018-097895-B-C41. GLP acknowledges financial support from Spanish Ministry of Science and Innovation through the Juan de la Cierva program (FJCI-2017-32370). JMM acknowledges the Spanish Ministry of Science, Innovation and Universities through FPU Program No. FPU18/01738.

# *Supporting Information*

# High power density energy harvesting devices based on the anomalous Nernst effect of Co/Pt magnetic multilayers

*Guillermo Lopez-Polin*[*,1], *Hugo Aramberri*[2], *Jorge Marques-Marchan*[1], *Benjamin I. Weintrub*[3], *Kirill I. Bolotin*[3], *Jorge I. Cerda*[1], *Agustina Asenjo*[*,1]

[1]Instituto de Ciencia de Materiales de Madrid (ICMM-CSIC), 28049, Madrid, Spain

[2]Materials Research and Technology Department, Luxembourg Institute of Science and Technology (LIST), L-4362, Luxembourg

[3]Department of Physics, Freie University Berlin 14195 Berlin, Germany

AUTHOR INFORMATION

**Corresponding Author**

* Guillermo Lopez-Polin. e-mail: guillermo.lopez-polin@uam.es

*Agustina Asenjo: e-mail: aasenjo@icmm.csic.es

## SI. 1 Device fabrication

### a) Macroscopic devices

For the deposition of the macroscopic devices, we used Si/SiNx substrates of 25x20mm$^2$. We used a window of 15x15mm$^2$ to grow a multilayer of this size centred on the substrate. The deposition was performed by sputtering in a HV chamber with a base pressure of 10$^{-6}$ mbar and deposition rates of ~0.1A/s. The heater and the heat sink were positioned in the top part of the substrates one at each side of the sample.

We prepared different multilayers with different thickness of Pt and Co. The hysteresis loops of the different multilayers have been measured by VSM (Vibrating-sample magnetometer) in two different configurations: applying perpendicular and parallel fields to the surface of the multilayer. Figure S1.1 shows the magnetic hysteresis curves of multilayers $Co_x/Pt_y$ with x= 0.8nm, 0.5nm and 0.25nm and y= 1.7nm, 1.5nm, 1.1nm, and 0.8nm. Only the multilayers with 0.5nm of Co and 1.1nm, 1.5nm and 1.7 nm of Pt show a remanence in the perpendicular direction of ~100%.

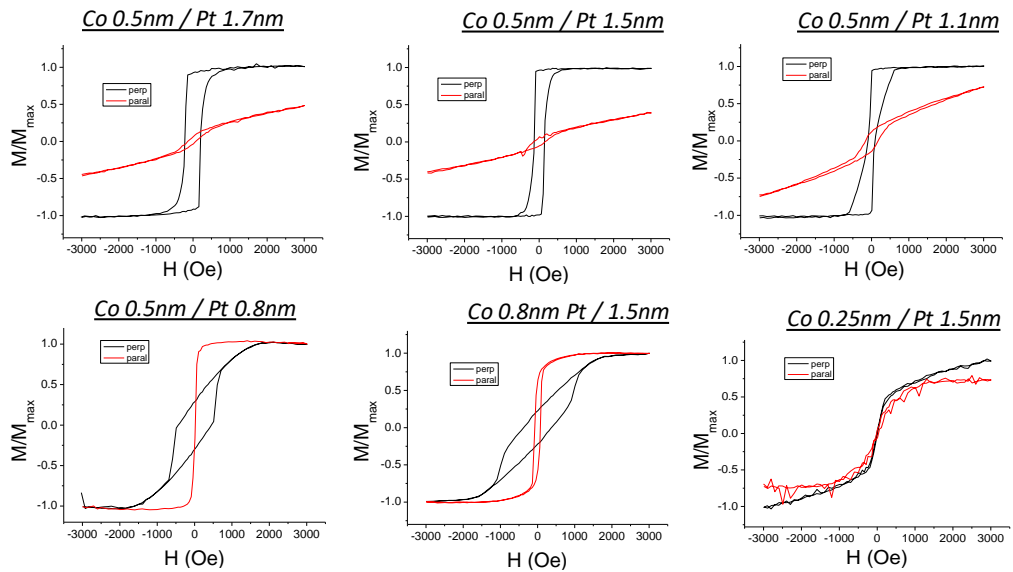

**Figure S1.1** *Hysteresis cycles of several multilayers with different thickness of Co and Pt.*

We measured the ANE response of the samples by inducing an in-plane thermal gradient and measuring the voltage in the in-plane perpendicular direction while applying a magnetic field perpendicular to the surface of the multilayer of ~280mT (above the saturation value). Figure S1.2 shows the lineal trend of the ANE voltage as a function of the temperature gradient across the sample for different multilayers (left) and the dependence of the ANE coefficient with the thickness of Pt and Co (right).

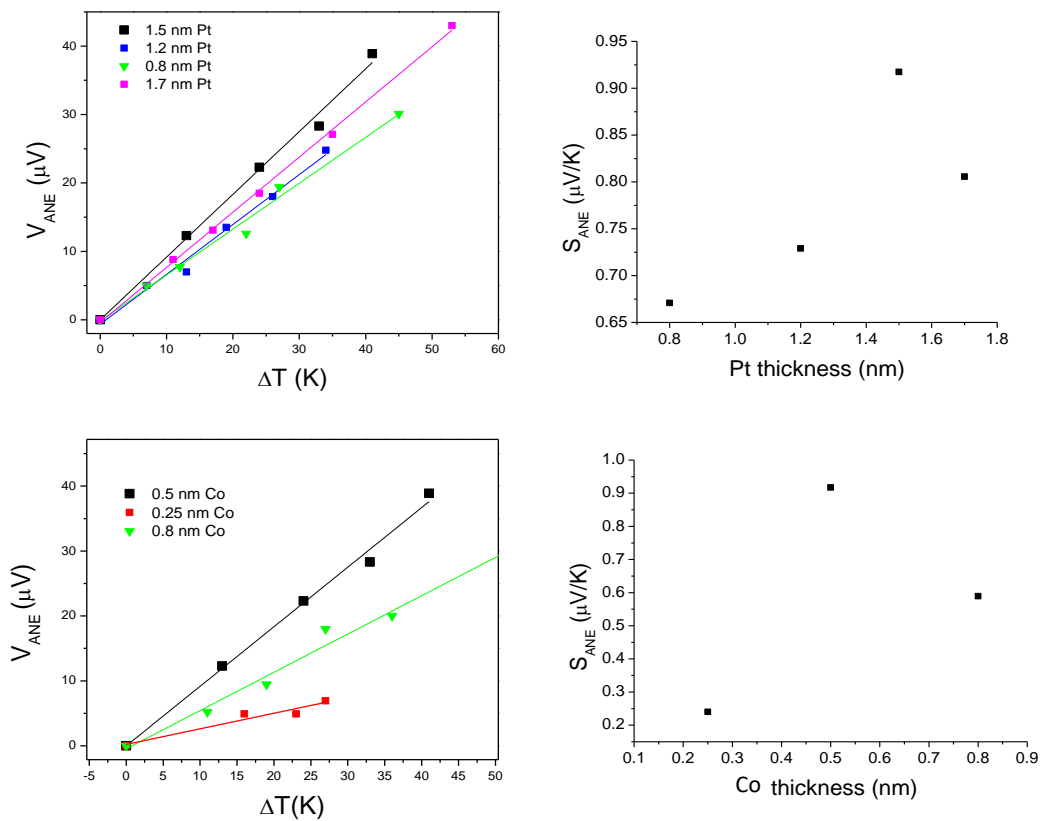

**Figure S1.2** *Dependence of the voltage obtained as a function of the temperature gradient in the transversal direction for various multilayers with different Co and Pt thicknesses. The maximum ANE coefficient was found for the multilayer with 0.5nm of Co and 1.5nm of Pt.*

### b) Atomic structure of the multilayers as measured by TEM

We performed TEM images of the multilayers in the "Servicio de Apoyo a la Investigación- *Servicio de microscopia electrónica de materiales*" in the university of Zaragoza. The goal was to obtain atomic resolution of the cross-section of the multilayer by TEM. To perform this experiment, it was necessary to obtain a lamella of the multilayer from the sample. To this end we deposited a layer of Pd on top of the surface to avoid electrostatic charges and we used a Focused Ion Beam (FIB) incorporated in a DUAL BEAM NOVA 200 system to cut a small piece with a thickness of ~25nm. The lamella was placed in an Image Corrected Titan TEM from FEI Company with a spherical aberration corrector (CEOS Company) at the objective lens. The images show the periodicity of the multilayers and also the atoms conforming the layers (figure S1.3). The roughness of the substrate induces a waving shape in the multilayer. We observed grains with uniform orientations of several nanometres size.

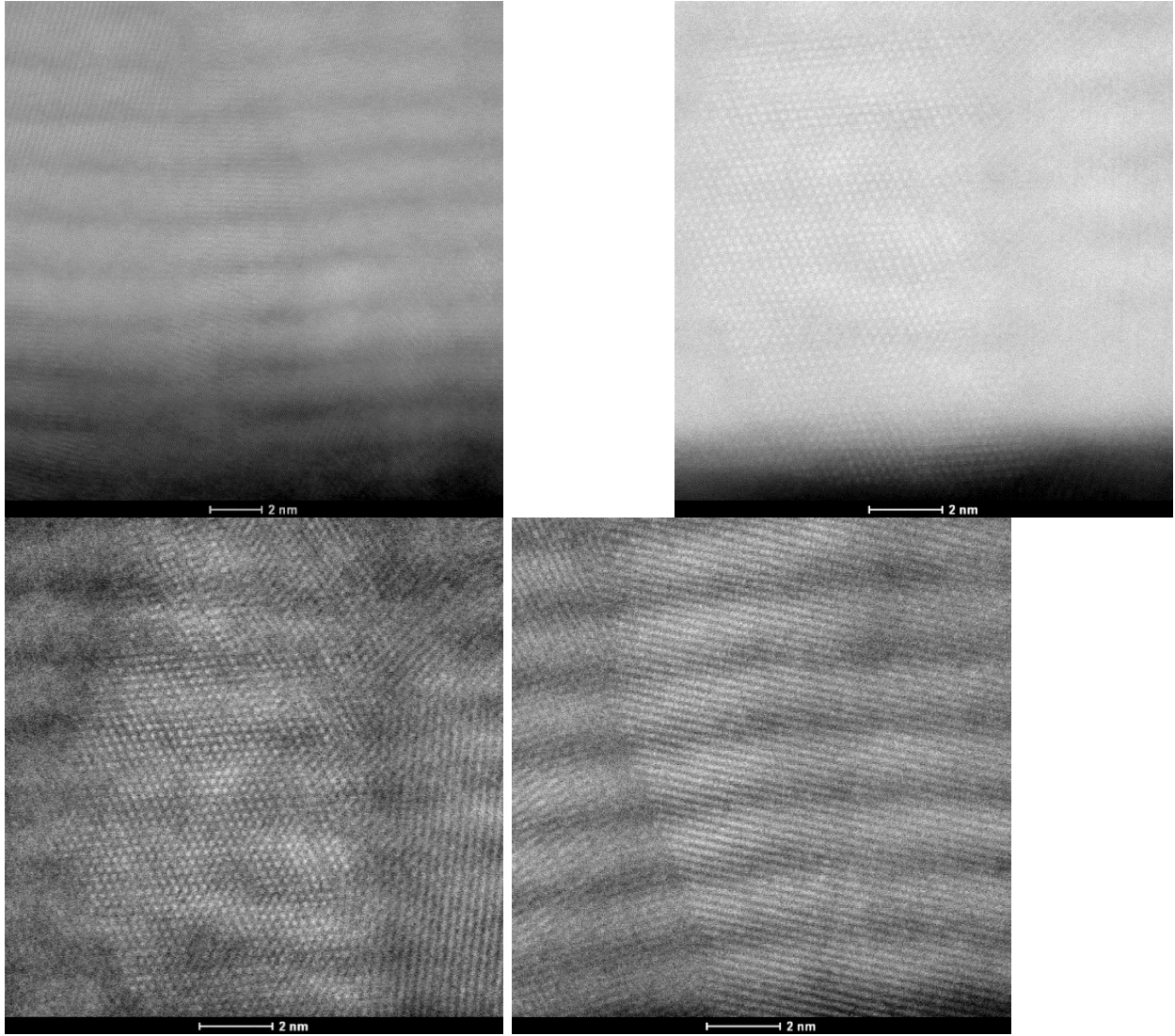

**Figure S1.3** *TEM images of the* [Co$_{0.5nm}$/Pt$_{1.5nm}$]$_{10}$ multilayers.

### c) Microscopic devices

Soda lime glass substrates, which have a poor thermal conductivity, have been used to prevent heat leakage out of plane. In addition, a thin layer of 50nm of alumina (Al$_2$O$_3$) was deposited on top of the glass to improve the in-plane thermal conductivity of the device. Rectangles of 100x5 µm$^2$ have been patterned by e-beam lithography to fabricate the multilayers bars. To increase the adhesion of the multilayers to the substrate, 1nm of Ti was thermally sublimated. After that, we deposited [Co/Pt]$_{10}$ multilayers by alternatively depositing 10 times 1.5 nm of Pt and 0.5nm of Co. The last layer of platinum was 2nm thick to better avoid oxidization of the cobalt. After that, we performed a second e-beam lithography for the heating and temperature sensing. Two 60nm thick Pt bars were deposited at a distance lower than 1µm, one at each side of the [Co/Pt]$_{10}$ rectangles. One of this bars acts as a heater and the other as a temperature sensor. For the temperature sensing, the resistance of the platinum bar was measured by using the 4 probes technique to avoid the contact resistances. The measurements of the ANE was performed by controlling the setup with the WSxM software [1] through a Dulcinea from Nanotec Electronics.

## SI. 2 COMSOL simulations

We performed finite element simulations using COMSOL Multiphysics[2]. COMSOL uses finite element modelling to compute the heat distribution. We used the Electromagnetic Heating interface, which includes the Electric Currents and Heat Transfer in Solids modules. The parameters for the materials used were given by the COMSOL library. Heat transfer coefficient of air (in contact with the top part of the device) was set to 5 W/(m²·K), and the thermal conductivity of alumina 2.7 W/(m·K) [3]. The electrical conductivity $\rho_e$ of Pt dependence of temperature $T$ was obtained from the calibration using $\rho_e = R \cdot A/L$, where R is the resistance, A the cross section and L the length between the inner contacts of the Pt sensor. The thermal conductivity $k$ of Pt was extracted from the electrical conductivity using the Wiedemann-Franz law $k \cdot \rho_e = L \cdot T$, where $L$ is the Lorenz number and is set to $2.6 \cdot 10^{-8}$ WΩ/deg² [4].

The thermal gradient in the x axis of the multilayer $\overline{\nabla T_x}$ and Pt probe temperature for different DC currents from 0 to 16 mA flowing through the microheater is evaluated. The geometry of the simulated device is shown in Figure S2, and the temperature distribution for a heater with a current $I_{heater}$ of 10 mA (density current of $2.38 \cdot 10^{10}$ A/m²) is shown in figure S2.b. Figure S2.c represents the mean temperature of the Pt probe between the inner contacts $T_{probe}$ as a function of the temperature gradient $\overline{\nabla T_x}$ averaged over the whole structure, which follows the equation $T_{probe}$ [K] $= (293.27 \pm 0.04)$ [K] $+ (12.561 \pm 0.011)$ [μm] $\cdot \overline{\nabla T_x}$ [K/μm]. The temperature gradients $\overline{\nabla T_y}$ and $\overline{\nabla T_z}$, in y and z axes respectively, in the multilayer are considered negligible for the further analysis of ANE effect. Moreover, since the $\overline{\nabla T_y}$ is 1000 smaller than the $\overline{\nabla T_x}$ the influence of the Seebeck effect is our measurements is negligible.

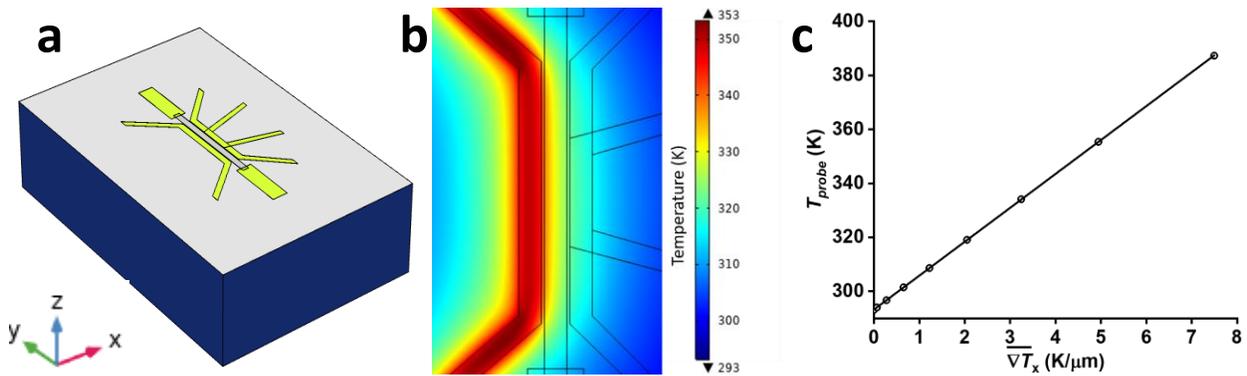

**Figure S2** *a) Device geometry used for computing the temperature distribution. b) Thermal distribution for a DC current through the microheater $I_{heater}$ of 10 mA. c) Simulated mean temperature of the Pt probe as a function of the thermal gradient in the structure along the x axis.*

## SI. 3 Calibration of the dependence of the resistivity of the platinum bar with temperature

The devices consist of a Pt bar that act as a heater, the Co/Pt stripe and a third bar with 4 contacts that acts as a thermal sensor. The resistivity of Pt has a well-known linear dependence with temperature, of approximately 3850 ppm/K near ambient temperature, and is widely used as a thermal sensor. However, thin films are well-known to have different resistivity vs. temperature dependence than the bulk material [5]. Thus, we experimentally determine the dependence of the resistance of the Pt sensor with temperature. We simultaneously measure the resistance of the sensor and the temperature of the sample with a thermocouple positioned close (few mm) to the sensor while heating homogeneously the whole sample (figure S3). The experimental ratio of the increase of the resistance with the temperature is 2700 ppm/K, much lower than the value for bulk Pt .

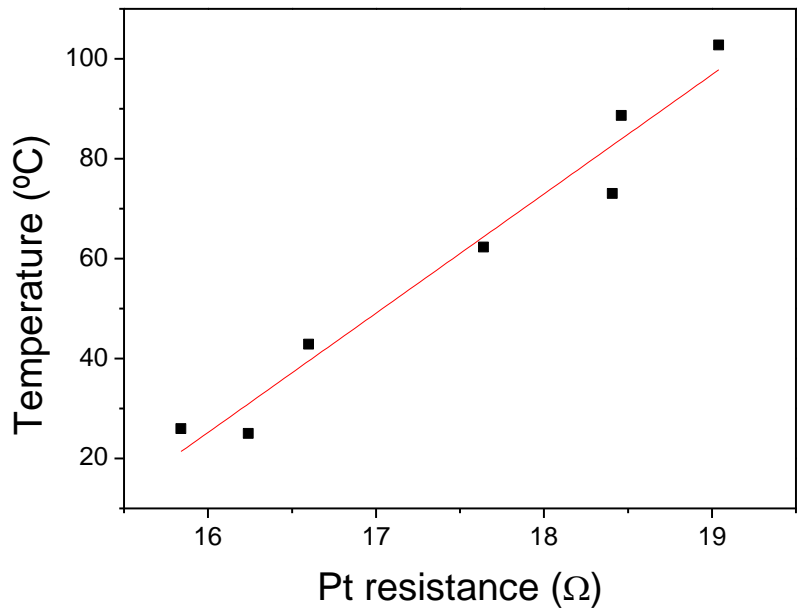

**Figure S3** Resistance vs Temperature data of the Pt thermal sensor.

## SI. 4 Maximum temperature gradient and ANE voltage in remanence

The remanence of the multilayers could be damaged when applying high temperatures. While at room temperature the structure is stable, when increasing the temperature domains in opposite direction appear in the multilayer and the magnetization of the sample tends to decrease. In our case, the maximum temperature gradient and ANE voltage that we could achieve without applying any field to the sample. The maximum Voltage was ~0.3mV and the maximum temperature difference across the multilayer was ~18K (~4K/µm).

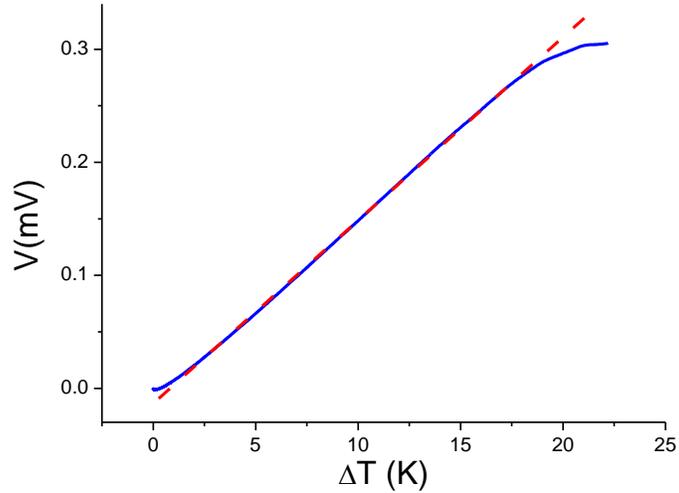

**Figure S4** $V_{ANE}$ *vs delta T curve obtained from a microscopic device of $[Co_{0.5nm}/Pt_{1.5nm}]_{10}$. At a thermal gradient of ~20K and a VANE of ~0.3mV the curve starts to deviate from the lineal tendency (dashed red line). We attribute the deviation to the appearance of domains in the opposite direction to the initial alignment due to thermal fluctuations. The deviation at low thermal gradients is due to the initial delay of the thermalisation of the system.*

## SI. 5 $V_{ANE}$ vs $\Delta T$ curves from experimental data

For the measurement of the Nernst effect of the microscopic devices we were able to directly determine 3 magnitudes: The voltage applied to the heating resistance, the resistance of the platinum thermal sensor and the transversal voltage generated due to ANE effect. From the resistance of the platinum bar we extracted the temperature of the sensor by calibrating the dependence of the resistance with temperature. From COMSOL simulations we related the average temperature of the sensor with the thermal gradient across the multilayer. At the end we were able to relate the thermal gradient across the Co/Pt structure with the resistance measured with 4 probe method.

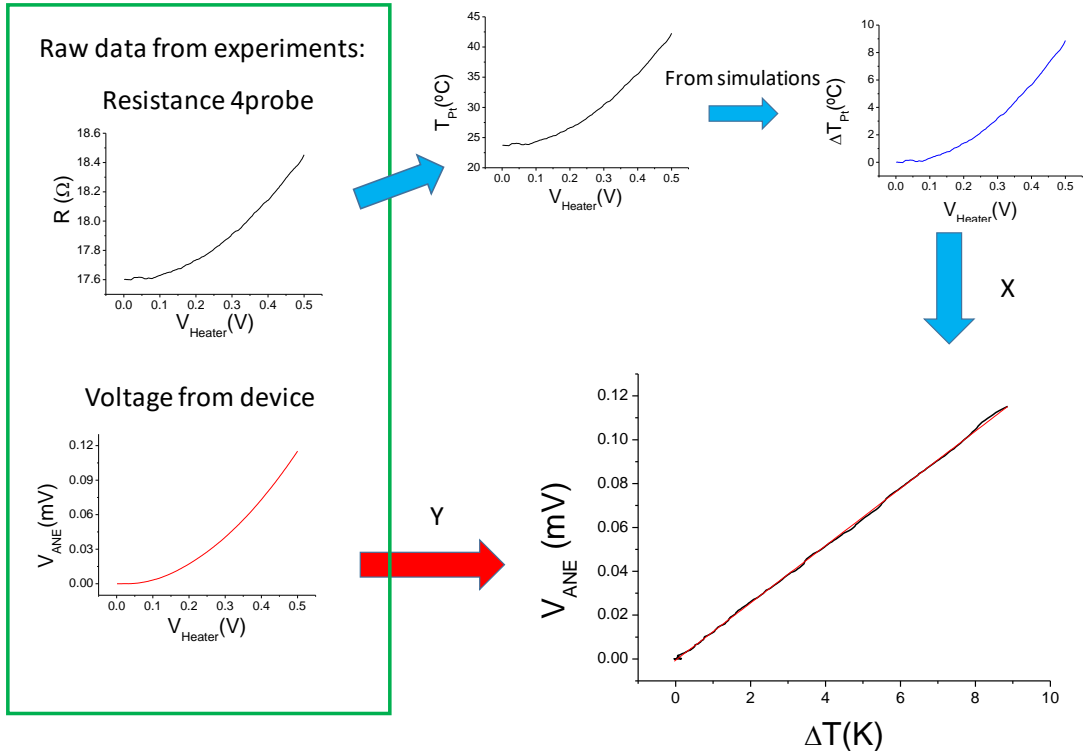

**Figure S5** *Scheme of the method used to extract the $V_{ANE}$ vs $\Delta T$ curves shown in the manuscript from the experiments.*

## SI. 6 MFM images of the microscopic devices

The magnetization of the sample is obtained from the MFM images measured simultaneously with the ANE voltage (figure S6). From the maps we can quantify the parts of the multilayer with the magnetization pointing in both directions, and therefore, the total magnetization of the sample. We processed the images with WSxM software. The histogram of the MFM images exhibit two overlapping Gaussians, each corresponding to the domains pointing upwards and downwards. We counted the number of points at the two sides of the intersection of the two Gaussians, which gives the number of points pointing to each direction. The corresponding volume is calculated to obtain the magnetization of the sample at each magnetic field [6].

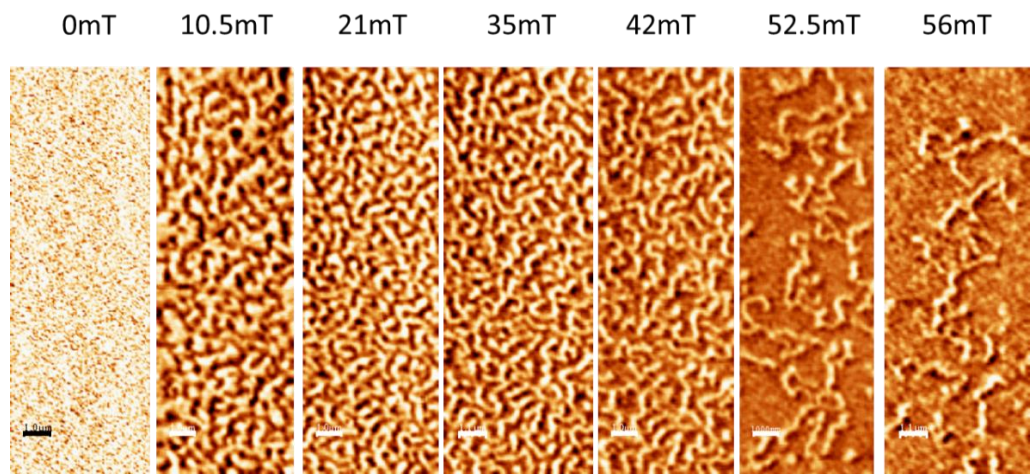

**Figure S6** *MFM images of the [Co$_{0.5nm}$/Pt$_{1.5nm}$]$_{10}$ multilayers as a function of the positive applied magnetic field. The initial state corresponds to the remanence after applied a saturating negative magnetic field.*

In order to measure the magnetic domains of the multilayers while minimizing the influence of the magnetic moment of the tip on the sample, we prepare homemade low-moment tips. For that we deposited a thin film of Co onto the front face of commercial Nanosensors AFM tips (PPP-FMR). We found that tips with a 12.5nm-thick layer of Co were just above the lower limit of sensitivity of the AFM to measure the magnetic signal of Co/Pt multilayers. These tips are used to determine the magnetization of the sample from MFM images (figures 3.b and S6). Otherwise, for the rest of MFM images (figure 1.e), we use standard commercial Nanosensors MFM tips (PPP-MFMR).

## SI. 7 First principles calculations

We obtain the atomic and electronic structure of the Co/Pt bilayers using first-principles density functional theory as implemented in the VASP package [7] [8]. To this end, we use the Perdew-Burke-Ernzerhof formulation of the generalized gradient approximation for the exchange-correlation functional [9]. We treat the atomic cores within the projector-augmented wave approach, considering the following states explicitly: 5$p$, 5$d$ and 6$s$ for Pt; 3$p$, 3$d$, 4$s$ for Co. We set the plane-wave cut-off to 500 eV, for which the results converge well. We employ Monkhorst-Pack k-point grids [10] of 7x7x2 for the 2/2 and 3/2 bilayers, and of 7x7x1 for the 4/2 and 5/2 systems. Each atomic layer is taken as hexagonally compact. Within the Pt layer the stacking pattern is *fcc* (*abcabc*), while in the Co layer the *hcp* stacking pattern is followed (*ABAB*), and the Pt is assumed to take the hollow site on top of the Co layer.

We allow the structures to relax until the atomic forces became smaller than 1 meV/Å and the residual stress becomes smaller than 0.01 GPa. The spin-orbit coupling (SOC) is included self-consistently using the second-variation method, employing the scalar-relativistic eigenfunctions of the valence states [11] as implemented in VASP.

We project the DFT-derived wave functions onto Wannier functions using the Wannier90 package [12].

We then use the resulting Wannier-based models to compute the Nernst conductivity $\alpha_{xy}$ using the Mott relation as follows [13]:

$$\alpha_{xy} = \frac{e}{T\hbar} \sum_n \int \Omega_{ij}^n(k)\{f(\varepsilon_{nk})(\varepsilon_{nk} - \mu) + k_B T \ln[1 + e^{-\beta(\varepsilon_{nk}-\mu)}]\}dk \qquad (3)$$

where $e$ is the electron charge, $T$ is the temperature, $\hbar$ is the reduced Planck's constant, $\Omega_{ij}^n$ is the Berry curvature of band $n$, $f$ is the Fermi-Dirac distribution function, $\varepsilon_{nk}$ is the band energy, $\mu$ is the chemical potential, $k_B$ is Boltzmann's constant and $\beta = (k_B T)^{-1}$. To this end we implement the calculation of equation (3) within the Wannier Tools package [14]. The Brillouin zone integral in $\alpha_{xy}$ is performed over 200x200x67, 200x200x51, 200x200x42, and 200x200x36 k-point grids for the 2/2, 3/2, 4/2 and 5/2 heterostructures respectively, which we found to be well-converged.

The search for topological band crossings is done using the built-in function of Wannier. We use a very fine k-mesh (with points every $7 \cdot 10^{-4}$ Å$^{-1}$) looking for band energy differences of less than 1 meV in the Wannier models for the heterostructures with SOC.

The results are shown in Figure S7. We see that thermoelectric conductivity oscillates with energy with an amplitude of a few Am$^{-1}$K$^{-1}$, peaking close to at E=+0.4 eV. However, at the computed Fermi energy the absolute values are not particularly large (except perhaps that of the 2/2 heterostructure). More importantly, $\alpha_{xy}$ at the Fermi level changes sign with the addition of each Pt monolayer, which is at odds with our experimental measurements. Even considering that the computed Fermi levels could be slightly off, in order to attribute a (dominant) topological origin of the ANE effect in the Co/Pt heterostructures there should be an energy window in which the computed thermopower does not change sign with Pt thickness and where the changes in the thermopower are not very large among the studied systems (the measured changes in ANE with Pt thickness are quite modest, of the order of 30 %, see SI2). This does not seem to be the case in the vicinity of the Fermi level, and could only be justified if the Fermi level were at +0.4 eV, but such a large error in the Fermi energy seems unlikely.

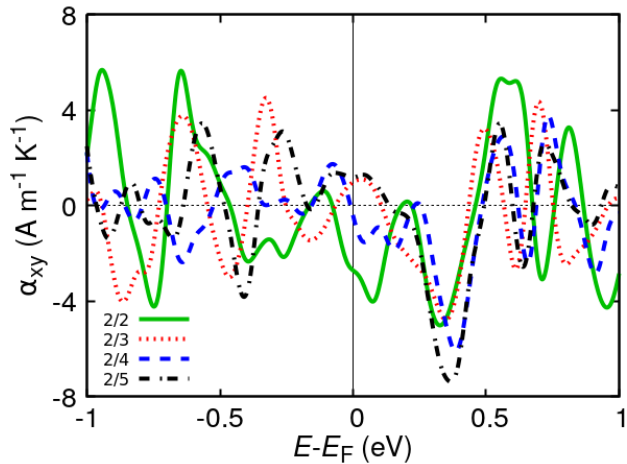

**Figure S7** *Room temperature thermoelectric coefficient $\alpha_{xy}$ as a function energy as computed from first principles for Co/Pt heterostructures. The results in solid green, dotted red, dashed blue and dot-dashed black correspond to a Pt thickness of 2, 3, 4 and 5 monolayers, respectively (see also legend). The Co is fixed to 2 monolayers in all cases.*

## SI. 8 Anomalous Hall effect measurements.

In order to extract information about the extrinsic or intrinsic origin of the ANE, we perform AHE measurements of the Co/Pt multilayers. To this end we grow a Co0.5nmPt1.5nm multilayer with H shape. We apply a current through the bar and measure the voltage between two of the contacts at the same side to determine the longitudinal resistivity ($\rho_{XX}$) and two contacts in opposite sides to determine the Hall resistivity ($\rho_{XY}$) of the multilayer.

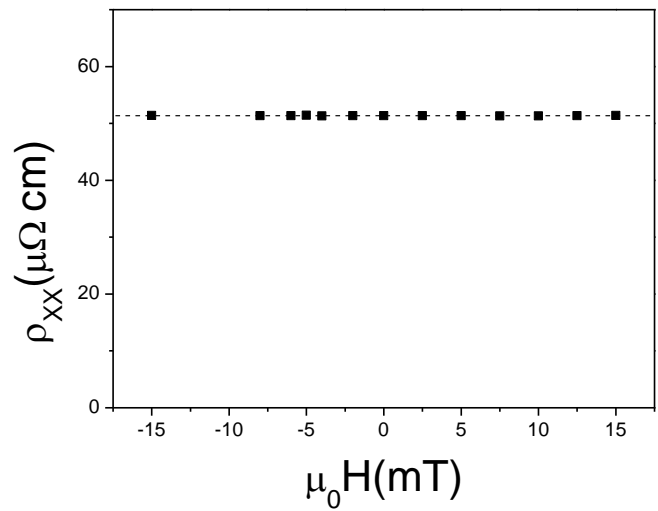

**Figure S8** *Longitudinal resistivity as a function of the magnetic field. The resistivity is constant with the field, showing no magnetoresistance. The dashed red line is place at a constant value of $\rho_{XX}$ to guide the eye.*

## SI. 9 Sensitivity of macroscopic devices

In order to test the sensitivity of the ANE voltage generated in the magnetic multilayers to the thermal gradient, we measured the voltage generated due to thermal energy caused by the human body by pressing with the finger one of the sides of the macroscopic [Co$_{0.5nm}$/Pt$_{1.5nm}$]$_{10}$ multilayer in remanence. Figure S8 shows the ANE voltage obtained from the device as a function of the time. During this time, we pressed and lift up three times the finger. The voltage due to the small temperature difference caused by the finger was small but measurable, of about 2μV.

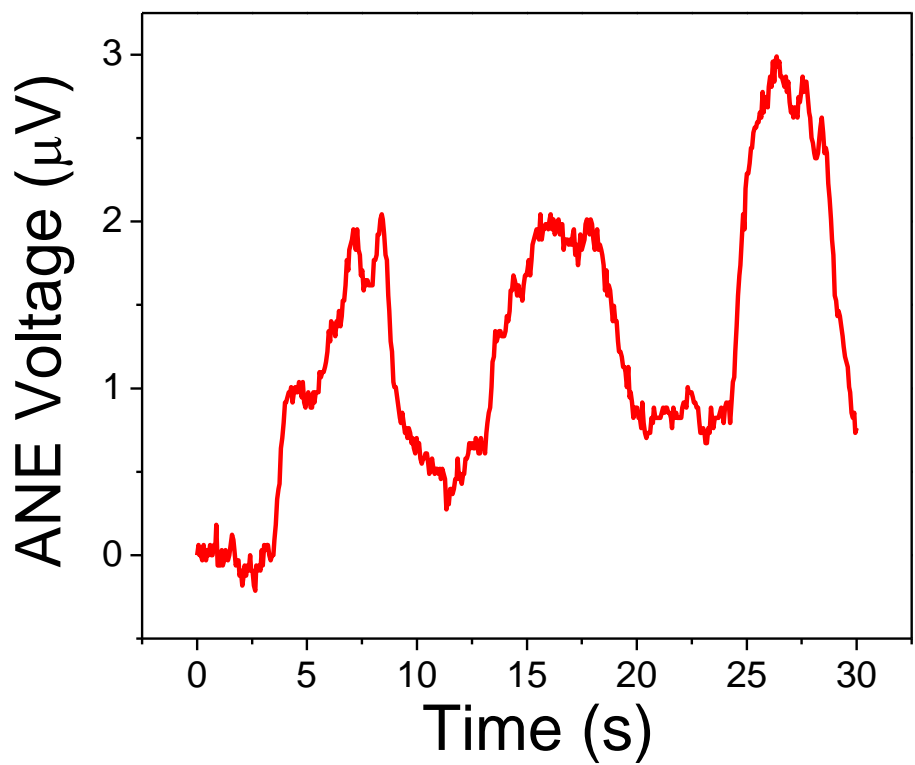

**Figure S9** *ANE voltage obtained when pressing with the finger one of the sides of the samples.*